# Mnemonics for Higher Education Using Contemporary Technologies

Dr.R.D.Balaji[1], Dr. Brijesh Ramniklal[1], N.Balasupramanian[1], Er. R.Malathi[2]

[1] Lecturer, Department of IT, Higher College or Technology, Oman,
[2] Sri Krishna Engineering College, Coimbatore, India

balajicas@yahoo.com[1], jajalbr@yahoo.com, balapondy111@yahoo.com,
bmalathisai@gmail.com[2]

## ABSTRACT

Education pierces into the contemporary world due to the influence of the emerging technologies in the digital world. Education field is also bombarded by the modern technologies and the demands of the digital society. It is affected by cloud computing, green computing and Internet. Still the success of this depends upon the minimization of the knowledge loss, while transferring ideas from teacher to student. Teaching & Learning is always an art. But when we take the traditional teaching and learning, knowledge loss is very high. As per our research, we have found that the knowledge loss is more than 60% for any subject on an average. Traditional Teaching& Learning is having advantage of more interaction, changing pace and technique based on audients. But it failed to address the knowledge loss due to mixed crowd, personal motivation and less possibility of the personal attention.

Even though modern technology is introduced along with the traditional method of teaching in Higher College of Technology (HCT), knowledge loss is very high still. Mnemonics is an old tool; still suitable for digital era students with the influence of recent technologies. Hence this paper discusses how to implement this mnemonics with the help of modern technologies like E-Learning and M-Learning. In this paper we have tried to discuss about E-learning and M-learning usage for implementing Mnemonics, and also discussed how far this concept can be used to make the Information and Communication Technology (ICT) more efficient and effective in term of making the students learning experience as a fun and without huge knowledge loss during the teaching and learning process including practical subjects. This paper also explains the initial research conducted with the students of database specialization of HCT and findings of that research about the knowledge loss. We have also suggests the future research areas in Mnemonics.

## INTRODUCTION

Education in the largest sense is any act or experience that has a formative effect on the mind, character or physical ability of an individual. In its technical sense, education is the process by which society deliberately transmits its accumulated knowledge, skills and values from one generation to another [3].

The evolution of culture, and human beings as a species depended on this practice of transmitting knowledge. In pre-literate societies this was achieved orally and through imitation. Story-telling continued from one generation to the next. Oral language developed into written symbols and letters. The depth and breadth of knowledge that could be preserved and passed soon increased exponentially. When cultures began to extend their knowledge beyond the basic skills of communicating, trading, gathering food, religious practices, etc., formal education and schooling, eventually followed [6]. The history of education is the history of man as since it's the main occupation of man to pass knowledge, skills and attitude from one generation to the other so is education.

Today, in all countries education has become the part and parcel of their development. Moreover, in recent days due to increased technologies students has diversified their thoughts towards recreation, as most of the modern innovations are exploitable. Hence, today students lack memory power and also they need arithmetic tools even for simple calculations. Hence, this paper is a tool for reintroducing a traditional tool called "Mnemonics" to help both the teachers and students of higher education to enhance their skills via improving memory which enlarges our world with the help of digital advanced technologies [11]. Without it, we would lack a sense of continuity and each morning encounter a stranger staring back from the mirror.

**EDUCATION AND TECHNOLOGY**

Technology is an increasingly influential factor in education. Computers and mobile phones are used in developed countries both to complement traditional education practices and develop new ways of learning [1]. This gives students the opportunity to choose what they are interested in learning. Technology offers powerful learning tools that demand new skills and understandings of students, including Multimedia, and provides new ways to engage students, such as Virtual learning environments. One such tool is virtual manipulative, which are an "interactive, Web-based visual representation of a dynamic object that presents opportunities for constructing mathematical knowledge" (Moyer, Bolyard, & Spikell, 2002). Emerging research into the effectiveness of virtual manipulative as a teaching tool have yielded promising results, suggesting comparable, and in many cases superior overall concept-teaching effectiveness compared to standard teaching methods. Technology is being used more not only for administrative duties in education but also in the instruction of students. The use of technologies such as power point and interactive white board is capturing the attention of students in the classroom. Technology is also being used in the assessment of students. Examples are the Audience Response system (ARS) and Moodle, which allows immediate feedback tests and classroom discussions [4].

Information and Communication Technologies (ICTs) are a "*diverse set of tools and resources used to communicate, create, disseminate, store, and manage information*" [8]. These technologies include computers, the Internet, broadcasting technologies (radio and television), and telephony. There is increasing interest in how computers and the Internet can improve education at all levels, in both formal and non-formal settings [7]. But still print remains the cheapest and most accessible therefore most dominant delivery mechanism in both developed and developing countries. In addition to classroom application and growth of e-learning opportunities for knowledge attainment, educators involved in student affairs programming have recognized the increasing importance of computer usage with data generation for and about students.

The use of computers and the Internet is in its infancy in developing countries, if these are used at all, due to limited infrastructure and the attendant high costs of access. Usually, various technologies are used in combination rather than as the sole delivery mechanism [8]. The Open University of the United Kingdom (UKOU), established in 1969 as the first educational institution in the world wholly dedicated to open and distance learning, still relies heavily on print-based materials supplemented by radio, television and, in recent years, online programming. Similarly, the Indira Gandhi National Open University in India combines the use of print, recorded audio and video, broadcast radio and television, and audio conferencing technologies [9]. The term "Computer-Assisted Learning" (CAL) has been increasingly used to describe the use of technology in teaching.

**EDUCATIONAL PSYCHOLOGY**

Educational Psychology is the study of how humans learn in educational settings, the effectiveness of educational interventions, the psychology of teaching, and the social psychology of schools as organizations. Educational psychology is concerned with the processes of educational attainment in the general population and in sub-populations such as gifted children and those with specific disabilities [8].

Educational psychology can in part be understood through its relationship with other disciplines. Educational psychology in turn informs a wide range of specialties within educational studies, including instructional design, educational technology, curriculum development, organizational learning, special education and class room management. Educational psychology both comes from and contributes to cognitive science and the learning sciences.

Sustainable capacity development requires complex interventions at the institutional, organizational and individual levels that could be based on some foundational principles [11]:

1. Strategies must be context relevant and context specific;
2. They should embrace an integrated set of complementary interventions, though implementation may need to proceed in steps;
3. Partners should commit to a long-term investment in capacity development, while working towards some short-term achievements;
4. Outside intervention should be conditional on an impact assessment of national capacities at various levels.

In some countries, there are uniform, over structured, inflexible centralized programs from a central agency that regulates all aspects of education [9].

- Due to globalization, increased pressure on students in curricular activities
- Removal of a certain percentage of students for improvisation of academics.

India is now developing technologies that will skip land based phone and internet lines. Instead, India launched EDUSAT, an educational satellite that can reach more of the country at a greatly reduced cost. There is also an initiative started by the OLPC foundation, a group out of MIT Media Lab and supported by several major corporations to develop a $100 laptop to deliver

educational software's. These will enable developing countries to give their children a digital education, and help close the digital divide across the world.

**Internationalization**

Education is becoming increasingly international. Not only are the materials becoming more influenced by the rich international environment, but exchanges among students at all levels are also playing an increasingly important role. Programs such as the International Baccalaureate have contributed to the internationalization of education. Some scholars argue that, regardless of whether one system is considered better or worse than another, experiencing a different way of education can often be considered to be the most important, enriching element of an international learning experience [4].

Educational research refers to a variety of methods, in which individuals evaluate different aspects of education including but not limited to: "student learning, teaching methods, teacher training, and classroom dynamics".

**Characteristics of Educational Research**

In his book entitled Fundamentals of Educational Research, Gary Anderson has outlined few characteristics that can be used to further understand what the field of educational research entails [4]:

• Educational research attempts to solve a problem.
• Research involves gathering new data from primary or first-hand sources or using existing data for a new purpose.
• Research is based upon observable experience or empirical evidence.
• Research generally employs carefully designed procedures and rigorous analysis.
• Research emphasizes the development of generalizations, principles or theories that will help in understanding, prediction and/or control.
• Research is carefully recorded and reported to other persons interested in the problem.

**Approaches in Educational Research**

There are two main approaches in educational research. The first is a basic approach. This approach is also referred to as an academic research approach. The second approach is applied research or a contract research approach. Both of these approaches have different purposes which influence the nature of the respective research [9].

Rudolf Steiner's model of child development inter-relates physical, emotional, cognitive, and moral development in developmental stages similar to those later described by Jean Piaget.

Developmental theories are sometimes presented not as shifts between qualitatively different stages, but as gradual increments on separate dimensions. People develop more sophisticated beliefs about knowledge as they gain in education and maturity.

**Motivation**

Motivation is an internal state that activates, guides and sustains behavior. Educational psychology research on motivation is concerned with the volition or will that students bring to a task, their level of interest and intrinsic motivation, the personally held goals that guide their behavior, and their belief about the causes of their success or failure.

**MNEMONICS**

A mnemonic is a strategy or device that helps us to store information in the long term memory and recall it when needed. A mnemonic device used by ancient Greek orators was the method of loci, or the location method, first described by Greek poet Simonides of Coes in 477 BC. This technique combines the principles of organization, visualization and association with something familiar, such as land mark on a road or an object in one's room or house.

An effective mnemonic for this is the acronym-combining the initial letter or letters of a group of words to form a new word or sentence. Many of us can easily remember the nine planets by an acronym/ acrostic- My Very Educated Mother Just Showed Us Nine Planets, where the beginning of each word of this acrostic gives us the hierarchy of the nine planets namely MERCURY, VENUS, EARTH, MARS, JUPITER, SATURN, URANUS, NEPTUNE AND PLUTO.

A mnemonic or mnemonic device is any learning technique that aids memory. Commonly met mnemonics are often verbal; something such as a very short poem or a special word used to help a person remember something, particularly lists, but may be visual, kinesthetic or auditory. Mnemonics rely on associations between easy-to-remember constructs which can be related back to the data that is to be remembered. This is based on the principle that the human mind much more easily remembers spatial, personal, surprising, sexual, humorous or otherwise meaningful information than arbitrary sequences.

**First letter mnemonics**

One common mnemonic for remembering lists consists of an easily remembered acronym, or phrase with an acronym that is associated with the list items. The idea lends itself well to memorizing hard-to-break passwords as well. For example, to remember the colors of the rainbow, use the mnemonic "Richard Of York Gave Battle In Vain".

**Assembly mnemonics**

In assembly language a mnemonic is a code, usually from 1 to 5 letters, that represents an Opcode, a number. Programming in machine code, by supplying the computer with the numbers of the operations it must perform, can be quite a burden, because for every operation the corresponding number must be looked up or remembered. Looking up all numbers takes a lot of time, and not remembering a number may introduce computer bugs. Therefore a set of mnemonics was devised. Each number was represented by an alphabetic code. So instead of entering the number corresponding to addition to add two numbers, one can enter "add". This type of mnemonic is different from the ones listed above in that instead of a way to make remembering numbers easier, it is a way to make remembering numbers unnecessary

# MNEMONICS IMPLEMENTATION

When we teach with traditional methodology the knowledge loss is more than 60 percentages in the practical subject like SQL Concepts. Hence we have made the materials with mnemonics. Simple commands are made with the help of assembly mnemonics and First letter mnemonics.

For example in the teaching of DDL commands create table command was made into mnemonics code like " CTTN (CNDTS);".the above code explains C for create, T for table, TN for table name then all the column specification must be enclosed in the round bracket. Within the bracket column name with data type and size to be specified and the columns should be separated with the comma. And all the commands must end with the semi colon. This made the students to remember this command very easily and able to reproduce whenever they want to write a DDL commands.

Similarly mnemonics code was created for the rules to use group by clause in the select students. It has three basic rules to avoid logical error in the command creation. Rule 1: Select column should be in Group clause. Rule 2: Select columns with group function no in Group Clause. Rule 3: Group columns no in Select Clause. The meanings of these three rules are given below. Rule 1: The columns specified which we are selecting by specifying in the select should be grouped by the group by clause. If some columns which are mentioned in the select statement, not specified in the group by clause will create error while executing the code. Similarly we have lengthy meaning for all the rules but when we make it short and teach them to the students, then it will be easy for them to remember these rules and implement it when they write command with group by clause. This avoids all the basic errors which may be done by the students during the practical exam.

Graphic mnemonics we have used to teach the set operators and join operators. To show the intersect we have used the overlapping two circles with the shade in the overlapped part.

Hence we have started involving section one students in the learning process with the rephrased materials with mnemonics and section two students with the normal and usual material. The section one students were thought and made them to use mnemonics codes to remember the commands and rules of the subject. Using this method student's knowledge loss level is very low. Also they remember the concept for a very long time. All the discussions which are related to their studies are also recorded and pass to their mobile devices. In the Mnemonics team the knowledge loss is almost zero. But other team knowledge loss percentage is higher than the mnemonics team. The advantage of M-learning is making the students to gain the knowledge in JIT also anywhere and anytime learning. We have to have a semantic web portal for implementing the mnemonics and personalized learning methodology. The web portal should be compatible with the mobile devices also. University can have a WI-FI facility throughout the campus, so that students can access the materials without any extra cost. Otherwise they have to depend upon the ISP to access their materials which will cost more to the students. When they have to pay extra money they are de motivated to use this technology. Hence M-learning and mnemonics are very much useful in the campus wide network, compared to the usage of this technology using ISP. The results absorbed after the final exam is recorded in the below given tables.

**Passing Statistics**:

|       | Section 1 | Section 2 |
|-------|-----------|-----------|
| Pass  | 25        | 27        |
| Fail  | 0         | 4         |
| Total | 25        | 31        |

Table 1: Pass statistics in both the sections of SQL Concepts

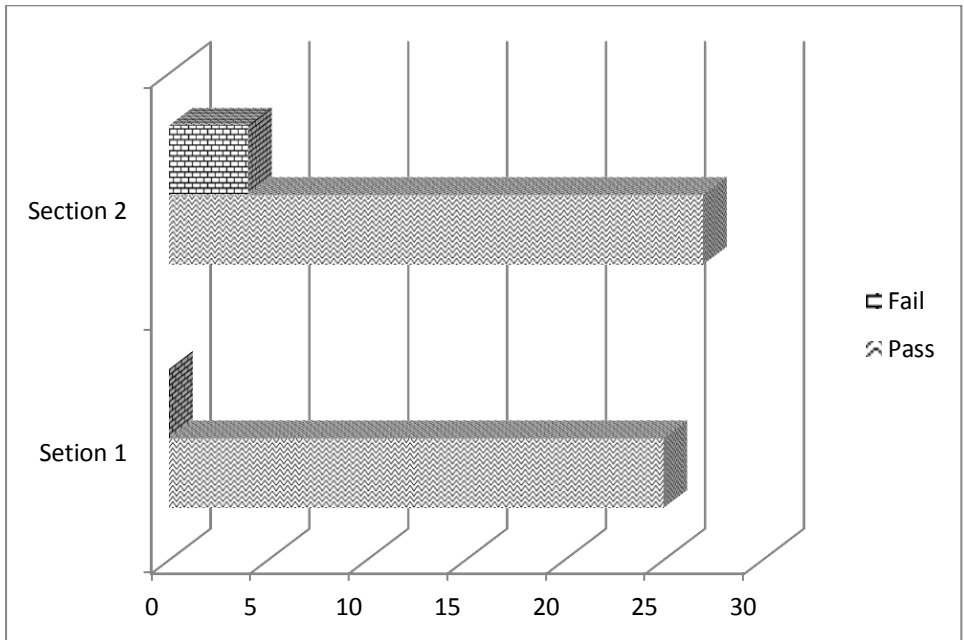

Figure 1: Bar Chart represent the pass percentage in both the sections of SQL Concepts

**Grade Statistics**:

| Grade | Section 1 | Section 2 |
|---|---|---|
| A | 10 | 4 |
| B | 8 | 11 |
| C | 7 | 14 |
| D(Fail) | 0 | 4 |

Table 2: Grade details of both the sections of SQL Concepts

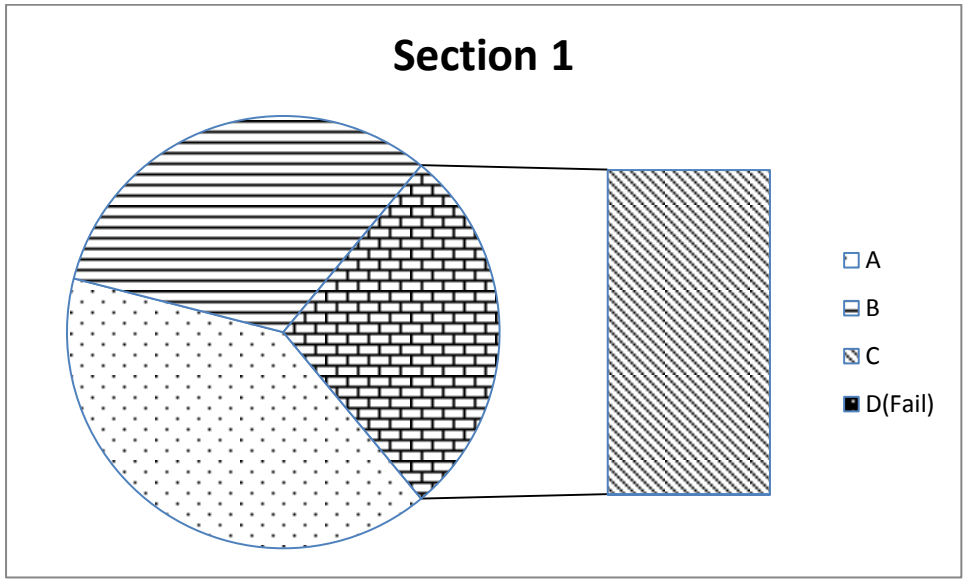

Figure 2: Section 1 grade distribution in SQL Concepts

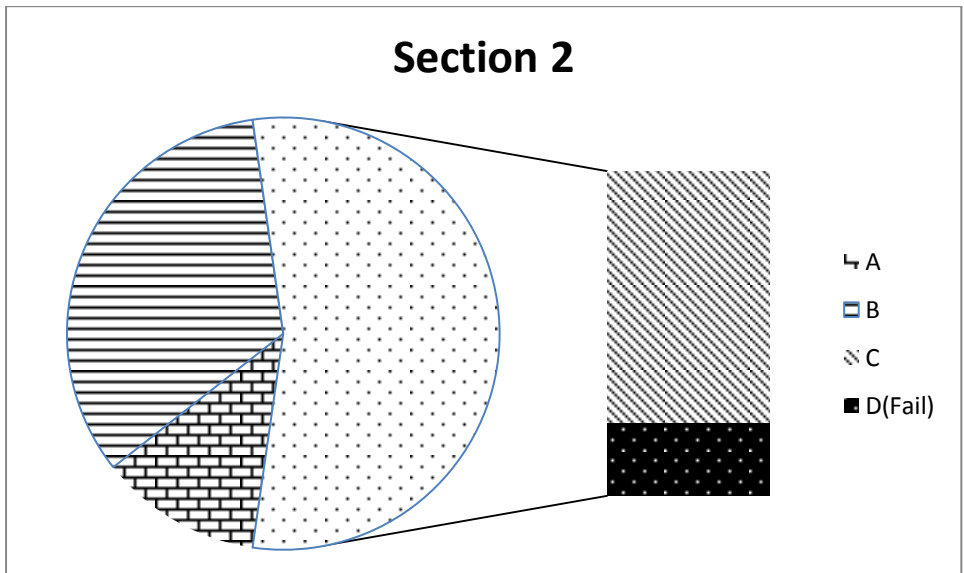

Figure 2: Section 2 grade distribution in SQL Concepts

It is very obvious that the mnemonics team performed well compared to the other team with the same material and same teacher. In mnemonics team also knowledge loss is there but the percentage is very low and due to many other factors.

**CONCLUSION**

All institutions need lot of new technologies and tools in the teaching and learning to avoid the knowledge loss. This paper discussed about mnemonics for memory improvisation and different methodologies to implement in the higher education field without much knowledge loss. Every technology and tool has its own economic and social impacts. M-learning is not given more importance in this paper since it is still in the beginning of research and implementation level. The implementation of M-Learning, methodologies, Requirements and Advantages are not discussed in this paper. But due to its contemporary nature we have taken the M-learning as a part of mnemonics implementation tool. This paper just discussed the introductory level of mnemonics in the practical paper with the advanced diploma students. The advantages, disadvantages, student's opinion and the consistency of this methodology can be evaluated only after the implementation of this technology for few more semesters. In future this may be introduced to the theoretical subjects and more practical oriented subjects, which needs creative thinking.